\documentclass[reprint, 10pt, floatfix,aps,twocolumn,a4paper,superscriptaddress,nofootinbib,notitlepage,pra,letterpaper]{revtex4-1}
\usepackage{inputenc}
\usepackage[T1]{fontenc} 
\usepackage{paralist}
\usepackage{hyperref}
\usepackage{amsopn,amsthm,dsfont,amssymb}
\usepackage{mathrsfs}
\usepackage{cancel} 
\usepackage{color} 
\usepackage{soul} 
\usepackage[normalem]{ulem}
\usepackage{cancel} 
\usepackage{hyperref} 
\usepackage{soul,xcolor}
\usepackage{lipsum}
\usepackage[english]{babel}

\hypersetup{
    colorlinks=true,
    linkcolor=blue,
    filecolor=magenta,      
    urlcolor=cyan,
}

\AtBeginDocument{%
    \newwrite\bibnotes
    \def\bibnotesext{Notes.bib}
    \immediate\openout\bibnotes=\jobname\bibnotesext
    \immediate\write\bibnotes{@CONTROL{REVTEX41Control}}
    \immediate\write\bibnotes{@CONTROL{%
    apsrev41Control,author="08",editor="1",pages="1",title="0",year="1"}}
     \if@filesw
     \immediate\write\@auxout{\string\citation{apsrev41Control}}%
    \fi
}%

\newcommand{\ket}[1]{|#1\rangle}
\newcommand{\bra}[1]{\langle #1|}

\usepackage{mathtools}

\usepackage{adjustbox}
\usepackage[caption=false]{subfig}

\begin{document}
\title{Detecting kHz gravitons from a neutron star merger with a multi-mode resonant mass detector}

\author{Germain Tobar}
\affiliation{Department of Physics, Stockholm University, SE-106 91 Stockholm, Sweden}

\author{Igor Pikovski}
\affiliation{Department of Physics, Stockholm University, SE-106 91 Stockholm, Sweden}
\affiliation{Department of Physics, Stevens Institute of Technology, Hoboken, New Jersey 07030, USA}

\author{Michael E. Tobar}
\affiliation{Quantum Technologies and Dark Matter Labs, Department of Physics,
University of Western Australia, 35 Stirling Hwy, 6009 Crawley, Western Australia.}

\begin{abstract}
We propose a multi-mode bar consisting of mass elements of decreasing size for the implementation of a gravitational version of the photo-electric effect through the stimulated absorption of up to kHz gravitons from a binary neutron star merger and post-merger. We find that the multi-mode detector has normal modes that retain the coupling strength to the gravitational wave of the largest mass-element, while only having an effective mass comparable to the mass of the smallest element. This allows the normal modes to have graviton absorption rates due to the tonne-scale largest mass, while the single graviton absorption process in the normal mode could be resolved through energy measurements of a mass-element in-principle smaller than pico-gram scale. We argue the feasibility of directly counting gravito-phonons in the bar through energy measurements of the end mass. This improves the transduction of the single-graviton signal, enhancing the feasibility of detecting single gravitons. 

\end{abstract}
\def\thefootnote{}\footnotetext{germain.tobar@fysik.su.se}
\maketitle
\section{Introduction} 
The direct detection of gravitational radiation has stimulated a rise in gravitational wave (GW) astronomy \cite{abbott2016observation,AbbottB.P.2016SotA,abbott2017gw170817,abbott2019narrow,abbott2023open}, with signals from both binary black hole (BBH) mergers and binary neutron star mergers (BNS). In addition, the capability to gain control over quantum systems at large mass scales is increasing rapidly, with the preparation of non-classical states of mechanical oscillators \cite{WollmanE.E.2015Qsom,ChuYiwen2017Qaws,SatzingerKJ2018Qcos,Ockeloen-KorppiCF2018Seom, PhysRevLett.130.133604}. This progress has driven new ideas for signatures of the quantum nature of gravity \cite{bose2023massivequantumsystemsinterfaces}, such as quantum features of gravity testing phenomenological modifications to known physics \cite{pikovski2012probing,MarinFrancesco2013Gbds,BushevP.A.2019Ttgu,CampbellWilliamM2023ICot}, as well as tests of collapse models \cite{RevModPhys.85.471,collapse1,KafriD2014Accm, belenchia2016testing,bassi2017gravitational,AltamiranoNatacha2018Gina,collapse2,ForstnerStefan2020Ntoq, collapse3, CarlessoMatteo2022Psaf, Tobar_2023_CSL}, quantum features of gravitational source masses \cite{bose2017spin, marletto2017gravitationally, PhysRevApplied.15.034004, pedernales2022enhancing, 2022fqce.book...85A, carney2022newton, PhysRevD.107.106018, higgins2024gravitationally}, and quantum features of gravitational radiation \cite{BlencoweMP2013Efta,PhysRevD.98.124006, parikh2020noise,kanno2021noise, Guerreiro2022quantumsignaturesin, AbrahãoLuca2024Tqoo}. In light of these advements it was recently proposed that single gravitons can be detected through projective measurements of the energy of a ground state cooled kg-scale resonant mass detector \cite{tobar2023detecting}, which would provide a gravitational version of the photo-electric effect. Implementing such a gravitational version of the photo-electric effect had previously been considered impossible \cite{DYSONFREEMAN2013IAGD}. 

While ground state cooling of a variety of mechanical oscillators has been shown to be feasible \cite{chan2011laser, DelićUroš2020Coal, barzanjeh2022optomechanics,rej2024nearground}, including the kg-scale LIGO mirror  \cite{whittle2021approaching} and tonne-scale bars \cite{4000phonons}, the most significant obstacle to the implementation of the graviton detection proposal of Ref.~\cite{tobar2023detecting} is performing direct measurements of single energy quanta of such a large mechanical oscillator, in particular phonons generated from sources of fundamental physics \cite{Tobar_2023_CSL}. The largest mechanical resonator this has been achieved for is currently on the order of micro-grams \cite{vonLüpkeUwe2022Pmit}. Furthermore, while the LIGO interferometers have been successful in detecting GWs mostly in their peak sensitivity range $\sim 100 \; \mathrm{Hz}$, their sensitivity decays at higher frequencies due to photon shot noise. While there are a range of detectors in use for GW searches at frequencies at $\mathrm{MHz}$ and above \cite{AggarwalNancy2021Caoo}, including the resonant mass schemes considered in this work \cite{GOrya2014,Tobarprl,CampbellWilliamM2023Tmag}, proposals have also been developed for GW detection at kHz frequencies for continuous signals with for example a superfluid resonant mass detector \cite{SinghS2017Dcgw}, or for sources from new physics with levitated optomechanics \cite{PRLhighfrequencyGeraci}. However, GWs at these higher frequencies are also expected from NS-NS mergers, for which numerical relativity simulations predict GW frequencies on the order of kHz\cite{SarinNikhil2021Teob}. Independently of implementing the gravito-phononic effect \cite{tobar2023detecting}, detecting a GW from a NS-NS merger at these higher frequency ranges can provide insights into post-merger remnants, yielding information on the hot nuclear equation of state \cite{SarinNikhil2021Teob, AckleyK.2020NSEM}. While interferometric detectors have also been proposed for observing the post-merger signals \cite{AckleyK.2020NSEM, BranchesiMarica2023SwtE,evans2023cosmicexplorersubmissionnsf}, resonant mass detectors may be suitable for targeting specific frequency bands for expected signals.

In this article, we propose the use of a multi-mode resonant bar \cite{Richard84,Price87,METobar1995} coupled to a parametric transducer \cite{MTobar95}, for the direct detection of up to kHz frequency gravitons emitted from a NS-NS merger. The signal is deposited into the either gram or pico-gram mass normal modes of the coupled oscillator system, while retaining the graviton absorption rate from the largest mass element (which can feasibly be on the order of tonnes \cite{blair94, 4000phonons, MCerdonio_1997, PhysRevD.76.102005}). A single graviton transition in the normal mode could then be resolved through a measurement of a single phonon transition in the smallest end mass (which as we show in this article, can in-principle have a mass lower than pico-grams). The substantially smaller end mass is suitable for optomechanical measurement of discrete single-phonon transitions. Furthermore, our proposal extends the possibility for resonant mass detection of gravitational radiation to a completely untested frequency regime for expected GW signals from BNS.

\begin{figure*}[t]
\begin{center}
\includegraphics[scale  = 0.55]{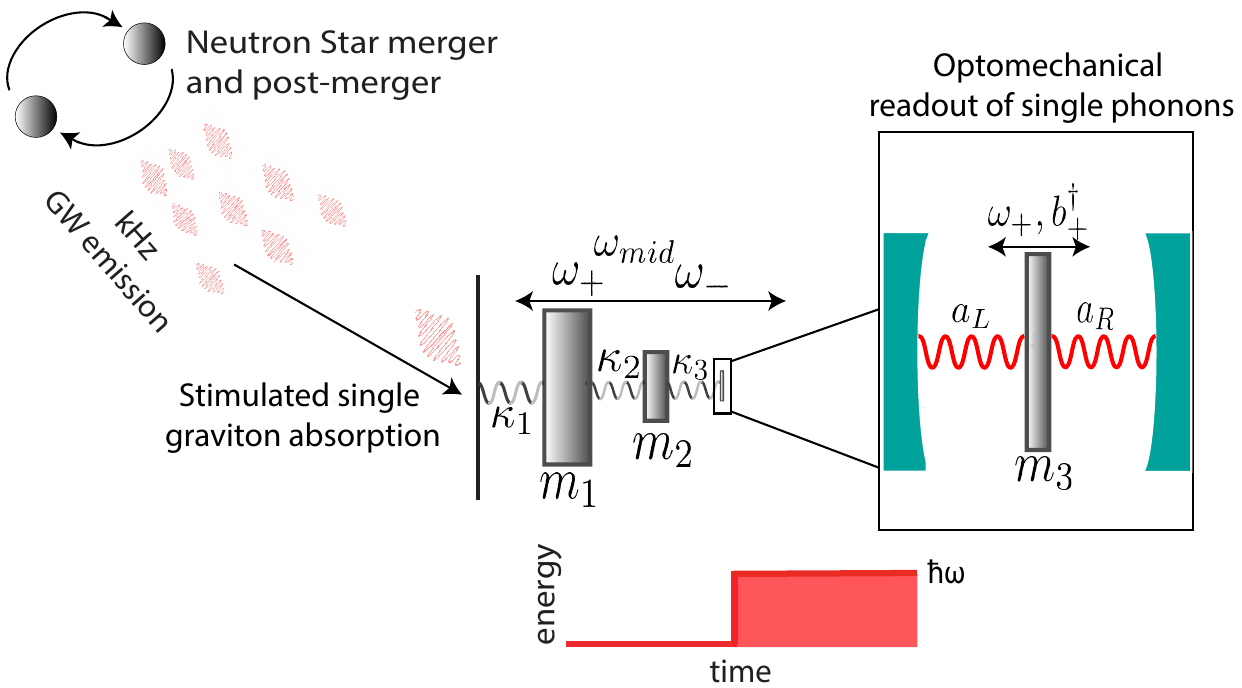}
    \caption{\label{absorption} kHz gravitational waves emitted from Neutron-Star mergers, result in stimulated absorption of gravitons in a multi-mode bar resonator. This can be used to implement a gravitational analogue of the photo-electric effect. Only the first mass is large enough to receive a non-negligible force from the passing gravitational wave, however, due to the strong coupling to the smaller masses, this will excite normal modes, causing graviton to phonon conversion into the normal mode of the coupled oscillator system. The phonon number of the normal mode can be measured through an optomechanical readout of the end mass. In this case, the problem of resolving a single graviton absorption, only requires the readout of the end mass, which can have lower than a pico-gram scale mass, while retaining the graviton absorption rate from the tonne-scale largest lumped mass. A measured single phonon transition in the normal mode, will correspond to a single-graviton-detection event. }
\end{center}
\end{figure*}

\section{Proposed protocol} 
The proposed set-up is shown schematically in Fig.~\ref{absorption}. We consider a series of $N$ strongly coupled mass elements of successively decreasing mass such that $m_{i+1} \ll m_{i}$, where $m_{i}$ is the effective mass of the i-th element. Our protocol involves firstly ground-state cooling each of the normal modes of the strongly coupled system, and subsequently performing measurements of single phonon transitions on the end mass induced by the gravitational wave. Here, we propose an optomechanical read-out, in which an electromagnetic cavity is coupled to the position of the smallest end mass. Due to the normal-mode splitting, coupling electromagnetic radiation to the end mass will allow for measuring single-phonon transitions in each of the normal-modes, which as previously explained, inherit the graviton absorption rate from the largest mass in the chain of coupled oscillators. Each of the normal modes can be read-out in rapid succession, providing an array of resonators that each have a comparable signal. If the mechanical mode is driven with a red-detuned pulse, an anti-stokes photon will only be produced when a single phonon is produced by the GW. Subsequent detection of the anti-stokes photon will herald a single phonon transition (and therefore a single graviton absorption from the GW). Correlation with a classical detection at LIGO can ensure the detector-click is due to the absorption of a graviton, rather than a noise phonon \cite{tobar2023detecting}.

\section{Single mode results} 
\begin{table}[h]
\begin{tabular}{lccc}
\hline \hline NS-merger  & \text { Inspiral } & \text { Merger } & \text { Post-merger } \\
\hline $\omega_0 / 2 \pi$ & $125 \; \mathrm{Hz}$ & $800 \; \mathrm{Hz}$ & $2410 \; \mathrm{Hz}$  \\
$h(\omega_0)$  & $2 \times 10^{-22}$ & $2 \times 10^{-22}$ & $5 \times 10^{-23}$  \\
\text {Material} & Sapphire & Sapphire & Sapphire \\
$v_s $& $10^4 \; \mathrm{ms}^{-1}$ & $10^4 \; \mathrm{ms}^{-1}$  & $10^4 \; \mathrm{ms}^{-1}$  \\
T & $10 \; \mathrm{mK}$ & $10 \; \mathrm{mK}$ & $10 \; \mathrm{mK}$\\
Q & $10^9$ & $10^9$  & $10^9$ \\
$M$ & $40 \; \mathrm{kg}$  & $15,000 \mathrm{kg}$ & $3500 \mathrm{kg}$ \\
\hline \hline
\end{tabular}
\caption{\label{tablesinglemode} Optimal mass for detecting single gravitons from a single-mode BAR resonator of resonance frequency $\omega_0$ in various frequency ranges of the gravitational waves emitted from a NS-merger at a distance of 40 Mpc, which corresponds to the distance at which the first and only GW NS-NS merger was observed \cite{abbott2017gw170817}.  The optimal mass is computed such that there is at least a $1 - \frac{1}{e}$ probability that a transition to an excited state will occur. The temperature and Q-factor is calculated such that the thermal-phonon rate integrated over a 0.5 s observation-window yields an excitation probability lower than $P \approx 0.6$. $v_s$ is the speed of sound of the proposed sapphire detector.}
\end{table}

Here we briefly review the dynamics of a single phononic mode of a bar interacting with (quantised) gravitational radiation considered initially in Ref.~\cite{tobar2023detecting}. The Hamiltonian for the bar as a mechanical oscillator of effective mass $M/2$ (where M is the total mass of the bar) and frequency $\omega_0$ interacting with quantised gravitational radiation is \cite{MaggioreMichele2007GwV1, tobar2023detecting}
\begin{equation}\label{H0}
\hat{H} = \hbar \omega_0 \hat{b}^{\dagger} \hat{b} + \frac{L}{\pi^2} \sqrt{\frac{M \hbar}{\omega_0}}\left(\hat{b}+\hat{b}^{\dagger}\right) \ddot{\hat{h}},
\end{equation}
where $\hat{h}$ is the quantised gravitational field perturbation in units of strain (which reduces to the classical metric perturbation $h$ in the semi-classical limit) and $\hat{b}^\dag$ ($\hat{b}$) are the creation (annihilation) operators for the bar's fundamental mode. In order to solve for the dynamics of a ground-state cooled bar interacting with gravitons, we adapt the solution for the time-evolution under Eq.~\eqref{H0} given in Ref.~\cite{tobar2023detecting}, for which the dynamics is captured by the unitary time-evolution operator $\hat{U}(t) =  e^{-i \omega_0 \hat{b}^{\dagger} \hat{b}} \hat{D}(\beta(t))$ with displacement operator $\hat{D}(\beta) = e^{\beta \hat{b}^{\dagger}-\beta^* \hat{b}}$, such that $\beta(t) = -i \frac{L}{\pi^2} \sqrt{\frac{M}{\hbar \omega_0}} \int_0^t d s \ddot{h}(s) e^{i \omega_0 s}$. Therefore, under the interaction with the gravitational wave, a ground-state-cooled bar evolves into the coherent state $|\psi(t)\rangle=\left|\beta(t) e^{-i \omega_0 t}\right\rangle$. In order to obtain at least a $1 - \frac{1}{e}$ probability that a transition to an excited state will occur (which quantum mechanically arises due to the absorption of a discrete number of gravitons), we solve for the mass of the bar such that $|\beta(t)| = 1$. The results are displayed in Table~\ref{tablesinglemode} for the interaction of a bar at three different resonance frequencies at which GW signals are expected from a neutron-star inspiral or post-merger signal. In the first column, we examine the response of a $125 \; \mathrm{Hz}$ bar mode interacting with a chirping GW signal which closely mimics the first NS-NS merger GW signal detected at LIGO \cite{abbott2017gw170817}. In the second and third column, we compute the mass which ensures $\beta(t) = 1$ using simulated results for the waveform prior to the merger (second column), and the gravitational wave's expected post-merger waveform (see Appendix~\ref{waveapp}). The simulated waveforms are taken from Ref.~\cite{SarinNikhil2021Teob}, such that it is the waveform expected from a NS-NS merger at a distance of 40 Mpc. This corresponds to within the range of the estimated distance at which the first and only GW signal NS-NS merger was observed \cite{abbott2017gw170817}, as well as within the range of the estimated masses for the binary components of the source of this signal. While at frequencies in the $\mathrm{hecto-Hz}$ range, this provides a known source of gravitational radiation, in the $\mathrm{kHz}$, post-merger signals are expected but unconfirmed. In this higher frequency range, resonance mass detectors could place constraints on the existence of the expected post-merger signal.

\section{Dynamics of a multi-mode bar interacting with gravitational radiation} 
Here we model the dynamics of a multi-mode resonant-mass detector interacting with gravitons. We model the system with position and momentum co-ordinates of each of the lumped elements as canonical co-ordinates  $[x_i, p_j] = \delta_{ij}$. The influence of the gravitons are modelled as a time-dependent external driving force only on the first lumped-element in the chain of coupled oscillators. After converting to the normal mode basis of the strongly coupled three-mode system, we can then expand the co-ordinate of the first-lumped mass in terms of the normal modes to derive the graviton absorption rate in each of the normal modes. Here, we present the 3-mode dynamics for conciseness, but the generalisation to the 5-mode dynamics is straight-forward (see Appendix~\ref{fivemodeapp}).

The Hamiltonian of a three-mode detector with a driving force of strength $g(t)$ input to the first lumped mass (Fig.~\ref{absorption}) is 

\begin{equation}
\begin{split}
    H=\frac{p_1^2}{2 m_1}+\frac{p_2^2}{2 m_2}+\frac{p_3^2}{2 m_3} &+ \frac{\kappa_1 x_1^2}{2}+\frac{\kappa_2\left(x_2-x_1\right)^2}{2} \\
    &+\frac{\kappa_3\left(x_3-x_2\right)^2}{2}+g(t) x_1,
\end{split}
\end{equation}
where $m_1, m_2$, $m_3$ are the effective masses of each of the individual elements and $\kappa_1, \kappa_2, \kappa_3$ are the spring constants. The assumption that only the first mass is driven by the gravitational wave is valid for $m_1 \gg m_{i > 1}$, since in this case the coupling strength of the subsequent smaller mass-elements to the gravitational wave is negligible. This can be seen through expressing the force on the first mass in terms of the gravitational radiation through  $g(t) = \frac{1}{\pi^2} L \sqrt{\frac{M}{\hbar \omega}} \ddot{h}(t)$ \cite{MaggioreMichele2007GwV1}, where ${h}(t)$ is the strain amplitude of the gravitational wave. If $m_1 \gg m_{i > 1}$, the coupling of the gravitational radiation to the smaller mass elements can be neglected.

Transduction to a nano-scale mass-element can be understood through conversion to the normal mode basis, to do this, we firstly convert to the following set of co-ordinates:
\begin{equation}
\vec{x}(t) \equiv\left(\begin{array}{l}
\sqrt{m_1} x_1(t) \\
\sqrt{m_2} x_2(t) \\
\sqrt{m_3} x_3(t),
\end{array}\right)
\end{equation}
in which case we can write the Hamiltonian as $H =  \frac{1}{2} \dot{\vec{x}}^T \dot{\vec{x}}+\frac{1}{2} \vec{x}^T \mathbf{M}_3 \vec{x}$,
where
\begin{equation}
\begin{aligned}
    \mathbf{M}_3 = \left(\begin{matrix}
\omega^2(1 + \eta) & -\omega^2\sqrt{\eta} & 0\\
-\omega^2\sqrt{\eta} & \omega^2(1 + \eta) & -\omega^2\sqrt{\eta}\\
0 & -\omega^2\sqrt{\eta}\ & \omega^2 
\end{matrix}\right)
\end{aligned}
\end{equation}
where we have assumed that the resonators are tuned (lumped element frequencies are all equal to be the value $\omega$, such that the spring constants are $\kappa_i = m_i\omega^2$), and that the ratios of successive mass elements are labelled as $\eta = \frac{m_{i+1}}{m_{i}} $. We will now proceed to diagonalise $\mathbf{M}$, such that $ \mathbf{D}_3 = \mathbf{P^T M_3 P}$ is a diagonal matrix  and the transformation $\vec{\varepsilon} = \mathbf{P}^T \vec{x}$ converts to the normal-mode co-ordinates, while noting that the vector $\vec{\varepsilon}$, is related to the normal-mode co-ordinates in the following way:
\begin{equation}
\vec{\varepsilon}(t) \equiv\left(\begin{array}{l}
\sqrt{m_+} x_+(t) \\
\sqrt{m_\mathrm{mid}} x_\mathrm{mid}(t) \\
\sqrt{m_-} x_-(t)
\end{array}\right)
\end{equation}
conversion to the normal mode basis in this way diagonalises the Hamiltonian of strongly interacting oscillators, allowing the full Hamiltonian to be expressed as
\begin{equation}\label{jjj}
H=\sum_{j=-}^{+}\left(\frac{p_j^2}{2 m_j}+\frac{m_j \omega_j^2 x_j^2}{2}\right)+ g(t)x_1 ,
\end{equation}
where $j \in -, \mathrm{mid},+$, labels the canonical co-ordinates, frequencies and effective masses of each of the normal modes. Under the above conditions of tuned resonators, and equal ratios of successive mass elements, there is a simple analytical expression for the normal mode frequencies
\begin{equation}
    \begin{split}
         \omega_+ =   \sqrt{(1+ \sqrt{ 2{ \eta }}+ { \eta } )} \omega \\
         \omega_\mathrm{mid} = \sqrt{(1 + { \eta } )} \omega \\
         \omega_- =   \sqrt{(1 - \sqrt{ 2{ \eta }}+ { \eta } )} \omega.
    \end{split}
\end{equation}

Now, expanding the canonical co-ordinate of the first lumped mass in the second term using $x _1 = P_{11} \sqrt{\frac{m_{\text {mid }}}{m_1}} x_{-} + P_{12} \sqrt{\frac{m_{-}}{m_1}} x_{\text {mid }} + P_{13} \sqrt{\frac{m_{+}}{m_1}} x_{+}$ (where $P_{ij}$ are is the $i,j$th element of the matrix $\mathbf{P}$), we find that when converted to the normal mode basis, the interaction term corresponds to a coupling of the gravitational radiation to each of the three normal modes. There is a simple analytical expression for the matrix $P_{ij}$, as given in Appendix~\ref{threemodegw1}, however, converting from the first lumped element to the three normal mode co-ordinates can be done with exact matrix elements $P_{11} = \frac{1}{2}, P_{12} = \frac{1}{\sqrt{2}}, P_{13} = \frac{1}{2}$. These factor of conversions to the normal mode co-ordinates being of order unity ensure that a signal can be deposited across each of the slightly detuned normal modes. Finally, quantising the normal mode co-ordinates and expanding in terms of creation and annihilation operators yields the following free and interaction Hamiltonian terms in the basis of the normal modes:
\begin{equation}\label{hminus}
    \hat{H} = \sum_{j=-}^{+} \hbar \omega_{j} \hat{b}_{j}^{\dagger} \hat{b}_{j}+\frac{P_{1j}}{\pi^2} L \ddot{h}(t) \sqrt{\frac{\hbar M}{ \omega_{j}}}\left(\hat{b}_{j}^{\dagger}+\hat{b}_{j}\right),
\end{equation}
and we see here explicitly that although only the first mass element is driven by the gravitational wave, a signal is deposited into each of the slightly detuned normal modes. 

\begin{table}[h]
\begin{tabular}{lccc}
\hline \hline  NS-merger  & \text { Inspiral } & \text { Merger } & \text { Post-merger } \\
\hline $\omega_{\mathrm{mid}} / 2 \pi$ & $125 \; \mathrm{Hz}$ & $800 \; \mathrm{Hz}$ & $2410 \; \mathrm{Hz}$  \\
$h(\omega_0)$  & $2 \times 10^{-22}$ & $2 \times 10^{-22}$ & $5 \times 10^{-23}$  \\
\text {Material} & Sapphire & Sapphire & Sapphire \\
$v_s $& $10^4 \; \mathrm{ms}^{-1}$ & $10^4 \; \mathrm{ms}^{-1}$  & $10^4 \; \mathrm{ms}^{-1}$  \\
T & $10 \; \mathrm{mK}$ & $10 \; \mathrm{mK}$ & $10 \; \mathrm{mK}$\\
Q & $10^9$ & $10^9$  & $10^9$ \\
$M_1$ & $40 \; \mathrm{kg}$  & $14,000 \mathrm{kg}$ & $4000 \mathrm{kg}$ \\
$M_3$ & $0.2 \; \mathrm{g}$  & $0.14 \; \mathrm{g}$ & $1 \; \mathrm{g}$ \\
$m_\mathrm{eff}$ & $0.5 \; \mathrm{g}$ & $0.6 \; \mathrm{g}$ & $6 \; \mathrm{g}$ \\
\hline \hline
\end{tabular}
\caption{\label{threemodetables} Optimal mass $M_1$ of the first lumped element in a three-mode-detector for optimising the probability of absorbing a single graviton from a NS merger signal in various frequency ranges. The optimal mass is computed such that there is approximately a $1 - \frac{1}{e}$ probability that a transition to an excited state will occur. The mass of the third and final element $M_3$, as well as the order of magnitude of the effective mass of the three normal modes $m_\mathrm{eff}$ are also displayed. The temperature and Q-factor is calculated such that the thermal-phonon rate integrated over a 0.5 s observation-window yields an excitation probability lower than $P \approx 0.6$. }
\end{table}
We now repeat the calculations performed in the single-mode case, but with the modified Hamiltonian for the coupling of the GW to each of the three normal modes given in Eq.~\eqref{hminus}. Extending to a five-mode example follows an analogous calculation to the three-mode example presented in the last section (see Appendix~\ref{fivemodeapp}). The results are displayed in Table~\ref{threemodetables} for a 3-mode example, and Table~\ref{fivemodetables} for a 5-mode example, which model the interaction of a bar at the same three different resonance frequencies at which GW signals are expected as considered in the earlier single mode example. We give detector parameters that allow for the single graviton absorption signal to be transduced to a gram-scale mass-element in Table~\ref{threemodetables}, and a picogram-scale mass-element in Table~\ref{fivemodetables} (such that successive measurements across each of the normal modes allow for approximately a $1 - \frac{1}{e}$ probability that a single-phonon-transition is recorded). 

While such multi-mode technology for gravitational wave detection is not new, only up to 2-3 mass elements were implemented in previous resonant mass detectors, with Niobe operating with a two-mode transducer \cite{blair94}, Auriga with a three mode transducer \cite{MCerdonio_1997,4000phonons}, the Schenberg antenna with a three mode transducer \cite{LiccardoV.2023Tdss}, the AGATA antenna with a three mode transducer \cite{BassanM.1994Ootn}, as well as the development of three-mode prototypes \cite{MarcheseLindaE.1994Poai}. Such detectors could be kept at cryogenic temperatures, with transduction from tonne-scale masses down to a mass of kg-scale, gram scale, or even milli-gram scale \cite{LiccardoV.2023Tdss}. Therefore, the technology to implement such a multi-mode resonant mass detector has been studied extensively, with substantial progress having been made towards the quantum ground state \cite{4000phonons}. However, since the advent of many of these experiments, even greater cooling power has been developed for large scale dark matter detection experiments \cite{PhysRevD.109.012009}, quantum computing, and tests of collapse models \cite{PhysRevLett.91.130401}. Therefore, it is within reach of existing technology to build kg-scale oscillators that are strongly coupled to smaller mass elements for transduction, ground state cooled normal modes, holding the device at $\mathrm{mK}$ temperatures as proposed in this work. The only significant modification that needs to be made from previous experiments is to supplement these existing detectors with modern phonon counting methods \cite{HongSungkun2017HBaT, VelezSantiagoTarrago2019PaDo, MirhosseiniMohammad2020Sqto,vonLüpkeUwe2022Pmit}, particularly for counting single phonons for testing fundamental physics \cite{Tobar_2023_CSL}, as we detail in Section.~\ref{readoutsec}.


\begin{table}[h]
\begin{tabular}{lccc}
\hline \hline  NS-merger  & \text { Inspiral } & \text { Merger } & \text { Post-merger } \\
\hline $\omega_{\mathrm{mid}} / 2 \pi$ & $125 \; \mathrm{Hz}$ & $800 \; \mathrm{Hz}$ & $2410 \; \mathrm{Hz}$  \\
$h(\omega_0)$  & $2 \times 10^{-22}$ & $2 \times 10^{-22}$ & $5 \times 10^{-23}$  \\
\text {Material} & Sapphire & Sapphire & Sapphire \\
$v_s $& $10^4 \; \mathrm{ms}^{-1}$ & $10^4 \; \mathrm{ms}^{-1}$  & $10^4 \; \mathrm{ms}^{-1}$  \\
T & $10 \; \mathrm{mK}$ & $10 \; \mathrm{mK}$ & $10 \; \mathrm{mK}$\\
Q & $10^9$ & $10^9$  & $10^9$ \\
$M_1$ & $40 \; \mathrm{kg}$  & $14,000 \mathrm{kg}$ & $4000 \mathrm{kg}$ \\
$M_5$ &  $0.26 \; \mathrm{pg}$  & $0.2 \; \mathrm{ng}$ & $500\; \mathrm{pg}$ \\
$m_{\mathrm{eff}}$ & $0.6 \; \mathrm{pg}$ & $0.8 \; \mathrm{ng}$ & $800 \; \mathrm{pg}$\\
\hline \hline
\end{tabular}
\caption{\label{fivemodetables}Optimal mass $M_1$ of the first lumped element in a five-mode-detector for optimising the probability of absorbing a single graviton from a NS merger signal in various frequency ranges. The optimal mass is computed such that there is approximately a $1 - \frac{1}{e}$ probability that a transition to an excited state will occur. The mass of the fifth and final element $M_5$, as well as the order of magnitude of the effective mass of the three normal modes $m_\mathrm{eff}$ are also displayed. The temperature and Q-factor is calculated such that the thermal-phonon rate integrated over a 0.5 s observation-window yields an excitation probability lower than $P \approx 0.6$.}
\end{table}

\subsection{Gravitational version of the photo-electric effect}
We now show the response of the detector discussed above is consistent with a single graviton exchange process, and summarise the analogy to the photoe-electric effect as originally presented in Ref.~\cite{tobar2023detecting}, but here extended to the normal modes of the oscillator. The interaction term of the gravitational radiation with each of the normal modes in Eq.~\eqref{jjj}, implies observing a discrete transition to an excited state in any of the normal modes involves the stimulated absorption of a discrete number of gravitons, and therefore will correspond to a gravitational version of the photo-electric effect \cite{tobar2023detecting}. To see this explicitly, we quantise the gravitational field interacting with the bar \cite{tobar2023detecting} by taking the mode expansion  $\hat{h}=\sum_{\mathbf{k}} h_{q, \mathbf{k}}\left(\hat{a}_{\mathbf{k}}+\hat{a}_{\mathbf{k}}^{\dagger}\right)$, where $h_{q, \mathbf{k}}=\frac{1}{c} \sqrt{\frac{8 \pi G \hbar}{V \nu_{\mathbf{k}}}}$ and $\hat{a}_{\mathbf{k}}\left(\hat{a}_{\mathbf{k}}^{\dagger}\right)$ are the annihilation (creation) operators for single gravitons with wavenumber $\mathbf{k}$ and frequency $\nu_{\mathbf{k}}$. Now, computing the transition rate of the resonator from the ground state due to the absorption of single gravitons from a coherent state of amplitude  $\alpha$ (assuming a monochromatic gravitational field, on-resonance with the $j$-th normal mode, and in the rotating-wave-approximation) using Fermi's golden rule \cite{ PhysRevD.109.096023, tobar2023detecting}, we obtain
\begin{equation}\label{abrate}
    \Gamma_{\text {stim }, j}  =\frac{ P_{1 j}^2 v_s^2}{4 \pi^3 \hbar} M h^2,
\end{equation}
as the graviton absorption rate in the $j$-th normal mode. Here, we have converted the coherent state amplitude to a strain amplitude through $|\alpha|^2 \rightarrow N$, where $N = \frac{h^2 c^5}{32 \pi G \hbar \nu^2}$ is the macroscopic number of gravitons in a gravitational wave of strain amplitude $h$ \cite{DYSONFREEMAN2013IAGD}. Estimating the absorption rate for a Niobium bar (with speed of sound $vs = 5000  \; \mathrm{ms^{-1}}$) of mass $M = 1800 \; \mathrm{kg}$, combined with a strain amplitude of $h = 5 \times 10^{-22}$, we obtain an absorption rate on the order of $1 \; \mathrm{Hz}$. This modifies the graviton absorption rate computed to be on the order of $1 \; \mathrm{Hz}$ for typical bar detectors in Ref.~\cite{tobar2023detecting} by the square of the factor of conversion $P_{1 j}$ between the first lumped element's co-ordinates and the normal mode co-ordinates (which is of order unity for tuned resonators as previously explained), further demonstrating that single graviton absorption processes can be significant for experimentally relevant normal modes of multi-mode resonant mass detectors. 

From Eq.~\eqref{abrate}, the archetypal signatures of the photo-electric effect are apparent. The threshold frequency manifests as the resonance condition - only when the gravitational wave frequency matches the detector's resonance frequency are gravito-phonons produced at the rate $\Gamma_{\text {stim }, j}$. The next signature is that the rate of particle production due to the absorption of gravitons is proportional to the square of the amplitude of the incident field (as the rate of the production of photo-electrons is proportional to the square of the electric field amplitude in the photo-electric effect). The next key signature is that the energy of produced phonon is independent of the intensity of the gravitational wave and proportional to its frequency. This can be summarised through a gravito-phononic analogue of the Einstein photo-electric relation for each of the normal modes: 
\begin{equation}\label{EPgravitons}
    \hbar \omega_j = \hbar \nu, 
\end{equation}
in that the energy of the produced gravito-phonon $(\hbar \omega_j)$ in the normal mode is proportional to the frequency of the incident gravitational radiation, while the classical energy density of the gravitational wave remains proportional to the square of the intensity of the incident radiation $E=\left(\frac{c^2}{32 \pi G}\right) \nu^2 h_0^2$. In analogy with early signatures of photons in the photo-electric effect \cite{einstein1905erzeugung}, this indicates that the field should consist of discrete $\hbar \nu$ packets of energy (gravitons) for a consistent explanation (an explanation which doesn't violate energy conservation). This is because the Einstein photo-electric relation in Eq.~\eqref{EPgravitons}, is an energy conservation relation - it indicates the discretisation of the field in units of $\hbar \nu$, for a consistent explanation of $\hbar \omega_j$ energy transitions in the detector.

Therefore, the observation of a discrete transition between energy eigenstates of a normal mode due to an incident gravitational wave, will also correspond to a signature of quantum gravity, in the same sense as we previously argued \cite{tobar2023detecting}, and to the same extent as the photo-electric effect for photons: although the transition can be explained semi-classically without quantising the gravitational field \cite{tobar2023detecting, PhysRevD.109.044009} (or the electromagnetic field in the photonic case \cite{lamb1968photoelectric, MandelLeonard1995Ocaq}), the semi-classical explanation violates energy conservation.

\section{Readout}\label{readoutsec}
We consider an optomechanical scheme \cite{RevModPhys.86.1391} for measuring the energy of the end mass $\hat{H}_{\mathrm{int}}=-\hbar G\hat{x}_3 \hat{a}_c^{\dagger} \hat{a}_c$, where the radiation pressure from an electromagnetic cavity with mode operators $\hat{a}_c$, $\hat{a}_c^{\dagger}$ couple to the position of the end lumped-element $\hat{x}_3$, with coupling strength $G$. Such an optomechanical read-out has been implemented in bars, using a capacitative coupling to a microwave cavity \cite{MTobar95}.  In order to see explicitly the enhancement in transductance due to the lower effective mass normal mode, we expand the canonical co-ordinate of the end lumped mass in terms of the normal mode co-ordinates as $\hat{x}_3 = P_{31} \sqrt{\frac{m_{\text {-}}}{m_3}} \hat{x}_{-} + P_{32} \sqrt{\frac{m_{\text {mid }}}{m_3}} \hat{x}_{\text {mid }} + P_{33} \sqrt{\frac{m_{+}}{m_3}} \hat{x}_{+}$. This provides an optomechanical interaction between the photonic cavity and each of the slightly detuned normal modes. However, in this case the single-photon optomechanical coupling strength to the $j$-th normal mode is $g_j = G P_{3j} \sqrt{\frac{\hbar}{2 m_{3} \omega_j}}$. This is a factor of $\xi = P_{3j}\sqrt{\frac{m_1\omega_1}{m_3\omega_j}}$ larger coupling strength than if the optical cavity is coupled directly to the first mass-element. For the mass-element parameters we consider in this work (see SI), we have $P_{3j} \sim 1$ and  $\sqrt{\frac{\omega_1}{\omega_j}} \sim 1$, such that $\xi \sim \sqrt{\frac{m_1}{m_3}}$. We have presented the three-mode example for simplicity, but this result generalises to the 5-mode example, providing an $\xi \sim \sqrt{\frac{m_1}{m_5}}$ larger coupling strength. 

 Transduction from a tonne-scale bar to a substantially smaller mass element has been achieved \cite{blair94}, with transducer technology developed in the resolved sideband regime \cite{CuthbertsonB.D.1996Pbei}, using smaller than kilo-gram or gram-scale elements for the end mass in such a set up will further enhance the single-photon optomechanical coupling strength. However, rather than measuring the individual quadratures of the microwave field, detecting single gravitons will require the modification that scattered microwave photons are registered on a photon-counter, to herald a single phonon transition (and therefore a single-graviton absorption).  
 
While the standard optomechanical interaction Hamiltonian doesn't directly couple to the oscillator's energy (as it is proportional to $\hat{x}$, rather than $\hat{x}^2$). Measurement of single-phonon transitions in mechanical oscillators from pure Fock states has been shown to be feasible even if there exists only a coupling to the position operator of the oscillator \cite{GallandChristophe2014Hsps}. If the oscillator is in the $n = 1$ Fock state, this can be resolved through the scattering and detection of an anti-stokes photon with frequency $\omega_c = \omega_j + \omega_L$ (where $\omega_L$ is the pump frequency, and $\omega_c$ is the cavity frequency). Detection of the scattered photon heralds the transition from the $n = 1$ Fock state. Measuring single-phonon transitions from pure Fock states in this way has been achieved experimentally \cite{HongSungkun2017HBaT, VelezSantiagoTarrago2019PaDo, MirhosseiniMohammad2020Sqto} in nano-mechanical oscillators in the resolved sideband regime, however with optical-readout at higher mechanical resonance frequencies. To model this explicitly, we consider the interaction between the mechanical oscillator and the electromagnetic cavity, which for a red-detuned pulse, simply becomes a beam-splitter interaction \cite{RevModPhys.86.1391}, $\hat{U}_\mathrm{int,j}(t) = \exp(-igt(\delta \hat{a}_c^{\dagger} \hat{b}_j+\delta \hat{a}_c \hat{b}_j^{\dagger}))$. In this case, if the cavity is initially in the vacuum state, the measurement operator on the mechanical oscillator, conditioned on the detection of a photon in the cavity is \cite{clarke2024nothingtheoreticalframeworkenhancing, cryerjenkins2024nothingenhancedlasercooling}
\begin{equation}
    \hat{M}_N = \bra{n}\hat{U}_\mathrm{int,j}(t)\ket{0} = \frac{\left(-i\sin(g\tau)\right)^N}{\sqrt{N!}} (\cos(g\tau))^{b^{\dagger}b} b^N.
\end{equation}
Given that the interaction with the gravitational wave drives the mechanics into the coherent state $\ket{\psi_m} = \mathrm{e}^{-|\beta|^2 / 2} \sum_{n=0}^{\infty} \frac{\beta^n}{\sqrt{n!}}|n\rangle_b$, after a waiting for the gravitational wave to produce this superposition state, a subsequent pulsed-optomechanical interaction of duration $\tau$, followed by the measurement outcome $\hat{M}_1$, produces the following state of mechanical subsystem:
\begin{equation}
\begin{split}
\ket{\psi_m} &= \\
     \mathrm{e}^{-|\beta|^2 / 2} &\sum_{n=1}^{\infty} \frac{\beta^n}{\sqrt{P(1)(n-1)!}}(-i\sin(g\tau)\cos(g\tau)^{n-1})|n-1\rangle,
\end{split}
\end{equation}
where $P(1) = \left\langle\psi_m\left|M_1^{\dagger} M_1\right| \psi_m\right\rangle$. In the regime for which $\beta < 1$, such a photon measurement approximately converts the mechanical state $\mathrm{e}^{-|\beta|^2/2}\ket{0} + \mathrm{e}^{-|\beta|^2/2}\beta\ket{1} + \mathcal{O}(\beta^2) \rightarrow \ket{0} + \mathcal{O}(\beta^2) $. Therefore, such a measurement of the optomechanical cavity in the $n = 1$ Fock state, corresponds to a measurement of the phonon number in the mechanical oscillator, signalling the transition from the $n = 1$ Fock state to the vacuum state. Furthermore, the probability for observing the measurement outcome $\hat{M}_1$ is zero if the mechanical oscillator remains in the ground state, and therefore, requires a single graviton absorption for this measurement outcome to be recorded. In contrast if no photon is recorded in the cavity, the relevant measurement operator on the mechanics $\hat{M}_0$ has unit probability for occurence if the mechanical oscillator is in the ground state. Therefore, a single graviton transition can be resolved, by applying a red-detuned pulse, and subsequent photo-detection of a single-photon with frequency $\omega_c$, corresponding to the measurement outcome $\hat{M}_1$ heralding a single-phonon transition. Assuming an input power of $1 \; \mathrm{mW}$ as achieved in Ref.~\cite{CuthbertsonB.D.1996Pbei} for a resonant mass detector's transducer, $\omega_j / 2 \pi  \sim \;1 \mathrm{kHz}$, with a single-photon optomechanical coupling strength of $g_0 / 2 \pi = 1 \; \mathrm{mHz}$ and cavity decay rate of $\kappa / 2 \pi \;\mathrm{Hz} = 275$, gives an optomechanical coupling strength of  $g / 2 \pi = 10^6 \; \mathrm{Hz}$. This places the system in the required regime of resolved sidebands, and sufficiently high optomechanical coupling strengths for a pulsed scheme to implement the measurement - pulsed time-scales on the order of $\mathrm{\mu s}$ are sufficient, followed by a series of filter cavities to filter the anti-stokes scattered photon from the pump as in Refs.~\cite{HongSungkun2017HBaT,GalinskiyI.2020Pcto}. For a micro-wave filter cavity with line-width $\mathrm{1} \; \mathrm{Hz}$, a series of filter cavities will be sufficient to filter an anti-stokes scattered photon from the $1 \; \mathrm{mW}$ pump, with $\kappa / 2 \pi \;\mathrm{Hz} = 275$ micro-wave cavity linewidth. We further note that such pulsed linear interactions followed by photo-detection can also be used for non-classical state preparation of the kg-scale oscillator \cite{GallandChristophe2014Hsps}.

Instead of using pulsed linear interactions supplemented by photo-detection to measure the energy, it is in-principle possible to witness a single-graviton absorption process through direct coupling of the photonic cavity to the square of the position operator. This can be achieved in an optomechanical setting through for example, a membrane-in-the-middle setup \cite{THOMPSONJ.D2008Sdco}, or magnetically levitated particles which have experimentally resolved a sensitivity to $x^2$ of the particle \cite{PhysRevApplied.19.054047}. An optomechanical interaction which achieves this coupling, can then again be expanded in terms of the normal modes, picking up contributions proportional to the square of the position operator of each of the normal modes. In this case, the coupling is enhanced by a factor $\frac{m_1}{m_5}$ compared to a direct coupling to the largest mass-element. This in principle reduces the difficulty of measuring a quantum jump between energy levels of a mechanical oscillator induced by the absorption of a single graviton to the simpler task of measuring a quantum jump in the orders of magnitude smaller end mass of the coupled-oscillator system. 

The possibility to measure the energy directly can be seen through the original Hamiltonian coupling the electromagnetic and phononic modes, $\hat{H} = \hbar \omega_{\text{cav}}(\hat{x}_3) \hat{a}^\dagger \hat{a} + \hbar \omega_{\text{m}} \hat{b}^\dagger \hat{b}$. By expanding the cavity frequency around an extremum where $\omega_{\mathrm{cav}}’(0) = 0$, the complete Hamiltonian becomes \cite{THOMPSONJ.D2008Sdco}:
\begin{equation}
    \hat{H}_{\mathrm{int}} = \hbar\left(\omega_{\text {cav }}(0) + \frac{1}{2} \omega_{\text {cav }}^{\prime \prime}(0) \hat{x}_3^2\right) \hat{a}^{\dagger} \hat{a} + \hbar \omega_{\mathrm{m}} \hat{b}^{\dagger} \hat{b},
\end{equation}
which shows a coupling to $\hat{x}_3^2$, and therefore to the energy of the end mass \cite{RevModPhys.86.1391}. By expanding the position of this mass in terms of the normal mode coordinates, interaction terms emerge that are proportional to the energy in each normal mode:
\begin{equation}
    \hat{H}_{\mathrm{int},j} = P_{3 j}^2\frac{1}{2} \omega_{\text {cav }}^{\prime \prime}(0) \hat{x}_j^2.   
\end{equation}
Consequently, if the coupling strength of the squared normal mode amplitude, $g_j^{(2)} = \frac{\hbar}{m_2\omega_j} \omega_{\text {cav }}^{\prime \prime}(0)$, surpasses that of the linear term, $g_j^{(1)} = \sqrt{\frac{\hbar}{m_2\omega_j}} \omega_{\text {cav }}^{\prime}(0)$, the dominant interaction with the microwave cavity will convey the oscillator’s energy. Therefore, if the end mass is placed at an extremum of the cavity mode profile a direct coupling to the energy of the normal can be achieved. This can be achieved if we consider for example, a $\mathrm{TEM}_{00}$ mode for the microwave cavity, featuring a sinusoidal mode profile along the cavity’s longitudinal axis (aligned with the motion of the mechanical oscillator). Following the framework from \cite{THOMPSONJ.D2008Sdco}, the cavity frequency is given by $\omega_{\text {cav }}(x) = \frac{c}{L} \cos^{-1}\left[\left|r_{\mathrm{c}}\right| \cos \left(\frac{4 \pi x}{\lambda}\right)\right]$, where  $r_c$  is the cavity’s reflectivity,  $\lambda$  the mode wavelength, and  $L$  the cavity length. For materials with high reflectivity (such as  $r_c = 0.995$ ), the cavity frequency profile,  $\omega_{\text{cav}}(x)$, peaks around  $x/\lambda = 0.25$ .

If the oscillator operates at frequencies in the kHz range and is in a thermal state at cryogenic temperatures around  $50 \; \mathrm{mK}$ , the amplitude uncertainty of the gram-scale membrane’s oscillations can be approximated as  $\sqrt{\mathrm{Tr}(\rho_\mathrm{th}\hat{x}_j^2)} = 10^{-12} \; \mathrm{m}$, where  $\rho_\mathrm{th}$  is the oscillator’s thermal state at these conditions. We further highlight that placing the resonator at a node of the micro-wave cavity profile, can be achieved with a split-post microwave resonator as undertaken recently for a high frequency resonator \cite{parashar2024upconversionphononmodesmicrowave}, replacing the BAW resonator with a smaller mass membrane at lower frequencies, and strongly coupling it to a kg-scale oscillator will enable the required energy measurement scheme suggested in this work. In order to enable QND measurements of the phonon number, the single-photon strong coupling regime $g_0/\kappa > 1$ will be required. Previously for a kg-scale Niobium membrane coupled to a $10 \; \mathrm{GHz}$ microwave cavity only $g_0/\kappa \sim 10^{-5}$ was achieved \cite{CuthbertsonB.D.1996Pbei}, however, extending the chain of coupled oscillators to nano-scale mass elements, could enable the single phonon strong coupling regime for a sufficiently small end mass. 


In this way, despite the largest system for which single phonon detection has been achieved being on the order of micro-grams \cite{vonLüpkeUwe2022Pmit}, as we have motivated in this section, there are multiple possibilities to explore for extending previously developed single-phonon detection methods for nano-scale masses \cite{vonLüpkeUwe2022Pmit,HongSungkun2017HBaT,GalinskiyI.2020Pcto} to larger masses, enabled through strong coupling to substantially smaller masses. While we have focused on the application of these techniques to the kg to tonne-scale resonant mass detectors considered in this work, the techniques studied here could also apply to the gram scale resonant mass detectors currently in active use for higher frequency gravitational wave detection \cite{GOrya2014,Tobarprl,CampbellWilliamM2023Tmag}. 

\section{Strain sensitivity for GW detection}
\begin{figure}[h] 
\begin{center}
\includegraphics[scale  = 0.48]{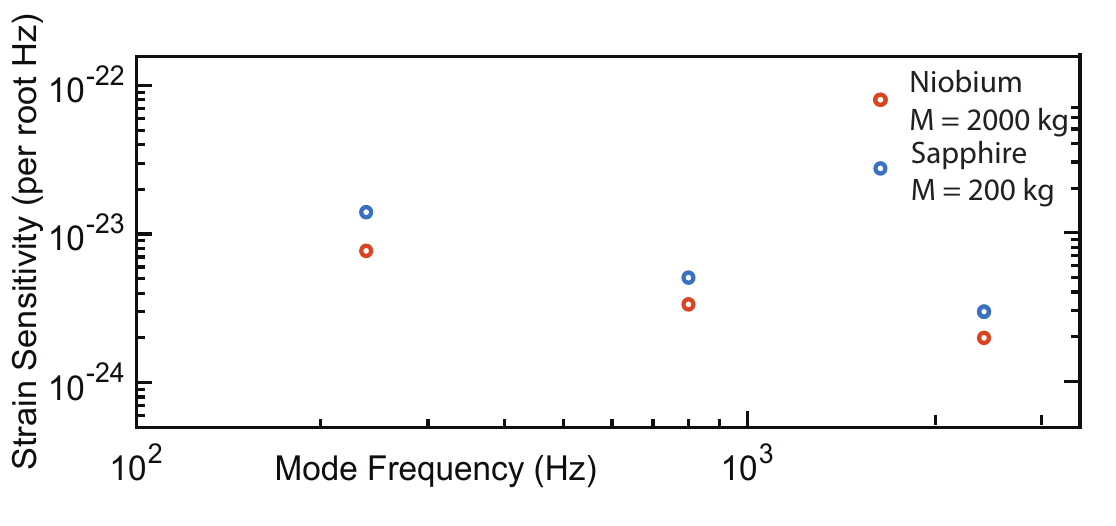}
    \caption{\label{strainh} The maximum spectral strain sensitivity of Niobium and Sapphire bars of total mass $M = 2000 \;  \mathrm{kg}$ and $M = 200 \; \mathrm{kg}$ respectively, optimised for three different frequencies. We assume either the Niobium or Sapphire bar has a Q-factor of $10^9$, and is in a thermal environment of temperature $T = 10 \; \mathrm{mK}$. At higher frequencies the peak sensitivity is up to an order of magnitude greater than the LIGO detectors, while the $125 \; \mathrm{Hz}$ bar modes have sensitivities comparable to that of LIGO's second observing run.}
\end{center}
\end{figure}

We now compute the spectral strain-sensitivity of the fundamental mode proposed bar detector. Assuming on-resonance interaction, the peak strain-sensitivity of the fundamental mode of the bar of resonance frequency $f = \omega/2\pi$ is \cite{MTobar95}
\begin{equation}
    h_{t h 1}^{+}(f)=\frac{\pi^2}{ L \omega} \sqrt{\frac{k_B T}{\omega Q m}},    
\end{equation}
where $L$ is the length of the bar, $Q$ is the Q-factor of the fundamental bar mode, $m$ its effective mass, and $T$ is the temperature of the thermal environment. The peak spectral strain-sensitivity for six different bars (made from either Sapphire or Niobium) that are optimised for the three different frequencies considered in previous sections is displayed in Fig.~\ref{strainh}.

The results displayed in Fig.~\ref{strainh} demonstrate that improved bar detectors, that operate at sub-kelvin temperatures, with high Q materials such as sapphire or Niobium, have peak sensitivities comparable to that of the LIGO detectors, even beating the sensitivity of the LIGO detectors at $2 \; \mathrm{kHz}$, by up to an order of magnitude. Despite their high-sensitivity on resonance, the main limitation of the bars in their existing form is their lower bandwidth. However, with existing interferometric detectors to cross-correlate detection events, the lower bandwidth does not limit the bar's capability to resolve single graviton-detection events (a gravito-phononic version of the photo-electric effect) if the bar's resonance frequency is placed within LIGO's detection band, as the detector's bandwidth can still be as large as $50-100 \; \mathrm{Hz}$ \cite{TobarMichaelE.2000PTft}. Furthermore, while the current interferometers can resolve the NS-NS signal with highest sensitivity in the inspiral phase, bar detectors can be placed at higher frequencies to obtain a higher sensitivity to the merger and post-merger stages than the current interferometric detectors.

These results demonstrate that while the current records for the bath temperature and Q-factor of resonant-mass detectors at frequencies on the order of $\sim 1 \; \mathrm{kHz}$ are $T \sim  100 \; \mathrm{mK}$ \cite{MCerdonio_1997,AllenZA2000Fsfg} and $Q \sim 10^8$  \cite{blair94} respectively, 1-2 orders of magnitude improvement in these parameters will allow for sensitivites competitive with the LIGO interferometers, and can be designed to target peak frequency ranges for which LIGO is not optimised. However, as also discussed earlier, even greater cooling power has been developed for modern experiments. Therefore, it is within reach of existing technology to construct kg-scale to tonne scale oscillators that are sensitive to known sources of gravitational radiation. Furthermore, Sapphire resonant mass detectors have achieved Q-factors as large as $Q \sim 10^{10}$ although restricted to lower masses and higher frequencies \cite{GOrya2014}. 

In recent years, resonant mass detectors have also been used for tests of fundamental physics beyond gravitational wave detection, such as for dark matter detection \cite{PhysRevLett.116.031102,PhysRevD.91.115023}, tests of lorentz invariance \cite{LoAnthony2016AToL,BushevP.A.2019Ttgu,CampbellWilliamM2021SfSD,CampbellWilliamM2023ICot}, as well as interferometric detectors. However, any such excitation would occur independently of gravitational wave signals detected by LIGO, given the lack of expected correlation between these phenomena. In this way, correlating detector clicks with the absorption of gravitational waves at LIGO, can provide independent confirmation, that the excitation was from the absorption of a graviton rather than any other multi-messenger source, and independent quantum sensor networks can provide evidence as to whether the excitation may be correlated with another multi-messenger source \cite{DaileyConner2021Qsna}. Even in the unlikely event such an excitation occurs from a dark matter source correlated with a LIGO detection event, the scaling in the excitation probability linearly with the mass predicted by the coupling to linearised gravity in Eq.~\eqref{hminus}, will enable confirmation across read-out of multiple resonators that the signal was from a source of gravitational radiation, rather than another source of new physics. Furthermore, while post-merger signals are expected at the kHz range and above, it is not yet ruled out that the presence of a viscous dark matter fluid, as indicated in Ref.~\cite{suárezfontanella2024gravitationalwaveemissionbinary} will place constraints on the possibility of observing such a signal. The presence of such a viscous dark matter fluid could reduce the detectable strain amplitude below the thermal noise threshold.

\section{Conclusion} 
In summary, we proposed a gravitational version of the photo-electric effect involving gravitons of up to KHz frequencies with a multi-mode bar. We have shown our proposed protocol gives realistic graviton absorption probabilities for experimentally relevant tonne-scale bars in three different frequency ranges at which GW emission is expected from NS-NS mergers.  Despite the largest mass being on the order of tonnes, the absorption of gravitons in this mass can be resolved by reading out single phonon transitions in the end mass which can be orders of magnitude less massive. The smaller effective mass of the coupled oscillator system provides orders of magnitude larger transductance as compared to the direct read-out of single phonon transitions of the largest mass-element. This enables reading-out a single graviton absorption process in a tonne-scale bar through measuring a single phonon transition in a mechanical oscillator, which is orders of magnitude smaller. In analogy with the photo-electric effect, recording a single graviton absorption process in the normal mode will provide the most compelling experimental evidence of the quantisation of gravity to date.  

{\bf Acknowledgments.} The authors would like to thank Vaishali Adya, Cyril Elouard, Sofia Qvarfort, Vivishek Sudhir, Witlef Wieczorek and Magdalena Zych for discussions. We thank Nikhil Sarin for providing the data from Ref.~\cite{SarinNikhil2021Teob} of the simulated NS-NS merger and post-merger signal. We thank Vivishek Sudhir for suggesting the idea to herald single graviton transitions through Raman scattering. MET was funded by the Australian Research Council Centre of Excellence for Engineered Quantum Systems, CE170100009 and Dark Matter Particle Physics, CE200100008. IP was supported by the National Science Foundation under Grant No. 2239498. GT acknowledges support from the General Sir John Monash Foundation.

\appendix

\section{NS-NS merger simulated signal} \label{waveapp}

In this work we have aimed to solve for the response of a single mode of a bar resonator to the higher frequency GW signals of an in-spiralling NS-NS merger than considered in Ref.~\cite{tobar2023detecting} - this is because the current LIGO detectors are most sensitive early in the inspiral phase. Once the emitted gravitational waves frequency chirp changes to a higher frequency, noise sources such as photon shot noise make the signal difficult to resolve. Therefore, we are interested in the higher-frequency-GW signal from a NS-NS merger, in particular the period in which the frequency of the emitted GW becomes as large as being on the order of 800 Hz, or the post-merger phase, where the frequency is expected to be on the order of 2.4 kHz. Detecting at these higher frequencies, would provide a GW detector optimised for an entirely new frequency range, while also lessening noise constraints, due to easier ground state cooling.

We now analyse the data from Ref.~\cite{SarinNikhil2021Teob}, of a simulated NS-NS merger: it is a simulation of a BNS with each neutron star having mass $1.35 \; \mathrm{M}_{\odot}$ and at a distance of 40 Mpc. However, we emphasize that this waveform is simulated assuming the BNS eventually forms a blackhole, and that there are other simulated waveforms which involve longer post-merger signals, we refer to Ref.~\cite{SarinNikhil2021Teob} for a comprehensive review.

\subsection{Pre-merger}
We start with the pre-merger signal, just prior to the merger event, which has a chirping waveform that chirps through the $\nu/2\pi \approx 800 \; \mathrm{Hz}$ frequency range. 

Now, the mass which optimises for single-graviton-exchange is,

\begin{equation}\label{chi111}
\begin{split}
    M &= \frac{\pi^2 \hbar \omega^3}{v_s^2 \chi(h, \omega, t)^2}, \\
    \chi(h, \omega, t) &= \left|\int_0^t d s \ddot{h}(s) e^{i \omega s}\right|. 
\end{split}
\end{equation}

Using the waveform from the simulated data, we obtain an optimal mass of 
\begin{equation}
    M \approx 14,791 \; \mathrm{kg}, 
\end{equation}
for a detector resonance frequency of $\omega/2\pi = 800 \; \mathrm{Hz}$ (and assuming a sapphire detector). For the pre-merger, the value of $\chi$, was evaluated to be: 
\begin{equation}
     \chi(h, \omega, t) \approx 9.5 \times 10^{-18}. 
\end{equation}

\subsection{Post-merger}
We now focus on the post-merger signal, which has a sinusoidal wave-form that has a frequency of $\nu / 2 \pi \approx 2410 \mathrm{~Hz}$.

We find that after calculating $\chi(h, \omega, t)$ for this waveform, and subsequently the optimal mass - that the response of the detector is approximately maximised if it's resonance frequency is $\omega/2\pi = 2410 \; \mathrm{Hz}$. We perform the calculations assuming $v_s = 10^4$ (assumes a sapphire detector), and find the following results of the calculation:

\begin{equation}
\begin{split}
    \chi(h, \omega, t) &\approx 1 \times 10^{-16} \\
    M &\approx 3524 \; \mathrm{kg}. 
\end{split}
\end{equation}

Therefore a sapphire detector of mass $M = 3524 \; \mathrm{kg}$, could be used to control the absorption of a single graviton from such a NS-NS post-merger signal. We now check the above calculation against an alternative estimate, which assumes the post-merger signal is a monochromatic wave of frequency $\omega/2\pi = 2410 \; \mathrm{Hz}$ and strain amplitude $h_0 \approx 7 \times 10^{-23}$, and find that it gives rough agreement with the above calculation for $\chi(h, \omega, t)$ - giving $\chi(h, \omega, t) \approx 7 \times 10^{-17}$ - this method involves the analytical estimate from Ref.~\cite{tobar2023detecting} of $\chi(\tau) \approx h_0 \frac{\omega^2 \tau}{2}$.

\section{Normal mode dynamics}
\subsection{Three mode} \label{threemodegw1}
Here, we consider the dynamics of three coupled oscillators, in which one of the oscillators is driven by a graviational wave. For previous studies of a gravitational wave interacting with a multi-mode detector we refer to Refs.~\cite{Richard84,Price87,METobar1995}, with optimisations of the number of modes described in Ref.~\cite{BASSANM1988Mrga}. The Hamiltonian is
\begin{equation}
    H = \sum_{i = 1}^3\frac{p_i^2}{2 m_i} + \frac{\kappa_1 x_1^2}{2} + \frac{\kappa_2 (x_2 - x_1)^2}{2} + \frac{\kappa_3 (x_3 - x_2)^2}{2} + g(t) x_1.
\end{equation}
The last term indicates a driving force on the first mass (which will be the gravitational wave, this coupling strength is $g(t) = \frac{1}{\pi^2} L \sqrt{\frac{M}{\hbar \omega}} \ddot{h}(t) $, where ${h}(t)$ is the strain amplitude of the gravitational wave). We now convert to the following set of co-ordinates:

\begin{equation}
\vec{x}(t) \equiv\left(\begin{array}{l}
\sqrt{m_1} x_1(t) \\
\sqrt{m_2} x_2(t) \\
\sqrt{m_3} x_3(t),
\end{array}\right)
\end{equation}

in which case, we can write the Hamiltonian as
\begin{equation}
   H =  \frac{1}{2} \dot{\vec{x}}^T \dot{\vec{x}}+\frac{1}{2} \vec{x}^T \mathbf{M}_3 \vec{x},
\end{equation}
where
\begin{equation}
\begin{aligned}
    \mathbf{M}_3 = \left(\begin{matrix}
\omega^2(1 + \eta) & -\omega^2\sqrt{\eta} & 0\\
-\omega^2\sqrt{\eta} & \omega^2(1 + \eta) & -\omega^2\sqrt{\eta}\\
0 & -\omega^2\sqrt{\eta}\ & \omega^2 
\end{matrix}\right)
\end{aligned}
\end{equation}
where we have assumed that the resonators are tuned (lumped element frequencies are all equal to be the value $\omega$, such that the spring constants are $\kappa_i = m_i\omega^2$), and that the ratios of successive mass elements are labelled as $\eta = \frac{m_{i+1}}{m_{i}} $. We will now proceed to diagonalise $\mathbf{M}$, such that $ \mathbf{D}_3 = \mathbf{P^T M_3 P}$ is a diagonal matrix  and the transformation $\vec{\varepsilon} = \mathbf{P}^T \vec{x}$ converts to the normal-mode co-ordinates, while noting that the vector $\vec{\varepsilon}$, is related to the normal-mode co-ordinates in the following way:
\begin{equation}
\vec{\varepsilon}(t) \equiv\left(\begin{array}{l}
\sqrt{m_+} x_+(t) \\
\sqrt{m_\mathrm{mid}} x_\mathrm{mid}(t) \\
\sqrt{m_-} x_-(t)
\end{array}\right)
\end{equation}
Computing $\vec{x}=\mathbf{P} \vec{\varepsilon}$, we take the example of a 3-mode system, where the lumped masses and frequencies are $\omega/2\pi = 125$, $\frac{m_{i+1}}{m_{i}} = 1/755$ (this was the ratio achieved in Niobe), then we can infer the following relationship between the position co-ordinate of the first lumped mass and the normal-mode position co-ordinates:
\begin{equation} \label{relation}
    \sqrt{m_1}x_1 \approx 0.48 \sqrt{m_-}  x_- + 0.72 \sqrt{m_\mathrm{mid}}  x_\mathrm{mid} + 0.51 \sqrt{m_{+}}  x_+.
\end{equation}
Therefore, the coefficient of conversion between the first lumped element position co-ordinate, and the normal mode co-ordinates is of order unity for these tuned lumped elements. In general, if we expand the position of the end mass $x_1$ in terms of the normal modes (as in the main text), we obtain:

\begin{equation}
    x _1 = P_{11} \sqrt{\frac{m_{\text {mid }}}{m_1}} x_{-} + P_{12} \sqrt{\frac{m_{-}}{m_1}} x_{\text {mid }} + P_{13} \sqrt{\frac{m_{+}}{m_1}} x_{+},
\end{equation} 
where $-, \mathrm{mid}, +$label the co-ordinates of each of the normal modes of increasing frequency. Furthermore, $P_{ij}$ denotes the (i,j) entry of $\mathbf{P}$.  Where the matrix $\mathbf{P}$, is given by
\begin{equation}
    \left(\begin{array}{ccc}
-\frac{1}{\sqrt{2}} & \frac{1}{2} & \frac{1}{2} \\
0 & \frac{1}{\sqrt{2}} & -\frac{1}{\sqrt{2}} \\
\frac{1}{\sqrt{2}} & \frac{1}{2} & \frac{1}{2}
\end{array}\right).
\end{equation}

We now write the full-Hamiltonian of the BAR interacting with the gravitational wave in the basis of the normal modes:
\begin{equation}
\hat{H}= \sum_{j=-}^+\hbar \omega_j \hat{b}_j^{\dagger} \hat{b}_j+\frac{1}{\pi^2} LM \ddot{h}(t)\hat{x_1},
\end{equation}
where 
$g = \frac{ML\ddot{h}(t)}{\pi^2} $ is the coupling strength. Now, when expanding $x_1$ in terms of the normal modes, and quantising the normal modes such that $x_j = \sqrt{\frac{\hbar}{2m_j\omega_j}}$ (here $m_j$ are the normal mode effective masses), and restricting to one of the normal modes (say the $+$ mode), we obtain the following total Hamiltonian (using that the effective mass of the first lumped element is $m_1 = M/2$):

\begin{equation}
  \hat{H} =  \sum_{j=-}^{+} \hbar \omega_{j} \hat{b}_{j}^{\dagger} \hat{b}_{j}+\frac{P_{1j}}{\pi^2} L \ddot{h}(t) \sqrt{\frac{\hbar M}{ \omega_{j}}}\left(\hat{b}_{j}^{\dagger}+\hat{b}_{j}\right).
\end{equation}

We can now repeat the calculations of Ref.~\cite{tobar2023detecting} but with coupling strength

\begin{equation}
    g_j(t)=\frac{P_{1j}}{\pi^2} \frac{\sqrt{M}L}{\sqrt{\hbar\omega_j}} \ddot{h}(t),
\end{equation}

and therefore, due to the interaction with the gravitational wave, the $j$-th normal mode of the BAR evolves into a coherent state of amplitude:

\begin{equation}\label{interactionterm}
\begin{split}
    |\beta| &= \frac{P_{1j}L}{\pi^2} \sqrt{\frac{M}{\omega_j\hbar}} \chi(h, \omega, t), \\
    &= \frac{P_{1j}v_s}{\pi \omega_1} \sqrt{\frac{M}{\omega_j\hbar}} \chi(h, \omega, t). 
\end{split}    
\end{equation}

where $\chi(h, \omega, t)$ is 

\begin{equation}
\chi(h, \omega, t)=\left|\int_0^t d s \ddot{h}(s) e^{i \omega_+ s}\right| .
\end{equation}

If we modify the first lumped-element effective mass to be as large as $14,000 \; \mathrm{kg}$, as well as the lumped-element frequencies to be as large as $\omega/2\pi = 800 \; \mathrm{Hz}$, using the input parameters to the multi-mode system as $m_1 = 7,000 \; \mathrm{kg}$, and $\eta = \frac{m_{i+1}}{m_i}=1/7000$. Using these parameters, we obtain normal mode frequencies $\omega_+ / 2 \pi = 807 \; \mathrm{Hz}$, $\omega_{mid} / 2 \pi = 800 \; \mathrm{Hz}$, $\omega_{-} / 2 \pi = 792 \; \mathrm{Hz}$, $\chi_+(h, \omega, t) \approx 9.26 \times 10^{-18}$, $\chi_-(h, \omega, t) \approx 9.26 \times 10^{-18}$, $\chi_{mid}(h, \omega, t) \approx 9.45 \times 10^{-18}$, we obtain coherent state amplitudes of (using Eq.\eqref{interactionterm}):

\begin{equation}
\begin{split}
    |\beta_+| = 0.49\\
    |\beta_{\mathrm{mid}}| = 0.70  \\
    |\beta_{\mathrm{-}}| = 0.50.
\end{split}
\end{equation}
In this case the combined expected phononic occupation number of the sum of all of the normal modes is approximately $1.0$. We now consider the case of the kHz frequency modes of the post-merger (which has a frequency of approximately $\omega / 2 \pi = 2410 \mathrm{~Hz}$). We use the following values for the lumped-element oscillator effective masses $m_1 = 2000\mathrm{~kg}$, with $\eta=\frac{m_{i+1}}{m_i}=1 / 2000$
(note that the mass of the largest mass is $M = 2m_1$ - as $m_1$ is the effective mass of largest mass as a simple harmonic oscillator), and lumped element frequencies $\omega/ 2 \pi  = 2410\mathrm{~Hz}$, i.e. we assumed that the resonators are tuned. We obtain the normal mode frequencies of $\omega_+ / 2 \pi = 2457 \; \mathrm{Hz}$, $\omega_{mid} / 2 \pi = 2410 \; \mathrm{Hz}$, $\omega_{-} / 2 \pi = 2364 \; \mathrm{Hz}$, this gives us the following wave-form parameters for each mode, $\chi_+(h, \omega, t) \approx 7\times 10^{-17}$, $\chi_-(h, \omega, t) \approx 5 \times 10^{-17}$, $\chi_{mid}(h, \omega, t) \approx 1 \times 10^{-16}$. This results in the following coherent state amplitudes:
\begin{equation}
\begin{split}
     \left|\beta_{+}\right|= 0.424 \\
    |\beta_{\mathrm{mid}}| =  0.61 \\
    |\beta_{\mathrm{-}}| = 0.3. 
\end{split}
\end{equation}
Combined this gives approximately a 50 percent chance that at least one of the three normal modes will undertake a detector click.

Next we consider the first lumped-element total mass to be $40 \; \mathrm{kg}$, as well as the lumped-element frequencjes to be $\omega/2\pi = 125 \; \mathrm{Hz}$, and mass-element input parameters $ m_1 = 20 \; \mathrm{kg}$ with $\eta=\frac{m_{i+1}}{m_i} = 0.001$. Using these parameters, we obtain normal mode frequencies $\omega_+ / 2 \pi = 128 \; \mathrm{Hz}$, $\omega_{mid} / 2 \pi = 125 \; \mathrm{Hz}$, $\omega_{-} / 2 \pi = 122 \; \mathrm{Hz}$, $\chi_+(h, \omega, t) \approx 1.13 \times 10^{-17}$, $\chi_-(h, \omega, t) \approx 1.13 \times 10^{-17}$, $\chi_{mid}(h, \omega, t) \approx 1.13 \times 10^{-17}$, we obtain a coherent state amplitude of (using Eq.\eqref{interactionterm}):

\begin{equation}
\begin{split}
    |\beta_+| = 0.503 \\
      |\beta_{\mathrm{mid}}| = 0.711  \\
    |\beta_{\mathrm{-}}| = 0.503.
\end{split}
\end{equation}

In this case the combined expected phononic occupation number of the sum of all of the normal modes is approximately $1.0$, giving a $1 - \frac{1}{e}$ probability that one of the normal modes will transition to an excited state due to the absorption of a single graviton.

\subsection{Five mode} \label{fivemodeapp}
\begin{equation}
\begin{split}
    H = &\sum_{i=1}^5 \frac{p_i^2}{2 m_i} + \frac{\kappa_1 x_1^2}{2} + \frac{\kappa_2 (x_2 - x_1)^2}{2} + \frac{\kappa_3 (x_3 - x_2)^2}{2} \\& + \frac{\kappa_4 (x_4 - x_3)^2}{2} + \frac{\kappa_5 (x_5 - x_4)^2}{2} + g(t) x_1.
\end{split}
\end{equation}
Repeating the procedure from Sec.~\ref{threemodegw1}, we convert to the following set of co-ordinates:

\begin{equation}
\vec{x}(t) \equiv\left(\begin{array}{l}
\sqrt{m_1} x_1(t) \\
\sqrt{m_2} x_2(t) \\
\sqrt{m_3} x_3(t)\\
\sqrt{m_4} x_4(t)\\
\sqrt{m_5} x_5(t)
\end{array}\right),
\end{equation}
and assume that the resonators are tuned, with identical ratios between the masses of successive elements $\eta=\frac{m_{i+1}}{m_i}=1$,such that the Hamiltonian can be expressed as
\begin{equation}
H=\frac{1}{2} \dot{\vec{x}}^T \dot{\vec{x}}+\frac{1}{2} \vec{x}^T \mathbf{M}_5 \vec{x},
\end{equation}
where 
\begin{equation}
\begin{aligned}
\mathbf{M}_5 = \left(\begin{matrix}
\omega^2(1 + \eta) & -\omega^2\sqrt{\eta} & 0 & 0 & 0\\
-\omega^2\sqrt{\eta} & \omega^2(1 + \eta ) & -\omega^2\sqrt{\eta} & 0 & 0 \\
0 & -\omega^2\sqrt{\eta}\ & \omega^2(1 + \eta)  & -\omega^2\sqrt{\eta}\ & 0 \\
0 & 0 & -\omega^2\sqrt{\eta}\  & \omega^2(1 + \eta) & -\omega^2\sqrt{\eta}\ \\
0 & 0 & 0 & -\omega^2\sqrt{\eta}\ & \omega^2
\end{matrix}\right).
\end{aligned}
\end{equation}
Once converted to the basis of normal modes, the same procedure can now be repeated from Sec.~\ref{threemodegw1}, to express the Hamiltonian as as a sum of the each of the normal modes driven by the gravitational wave:

\begin{equation}
\hat{H}=\sum_{j=--}^{++} \hbar \omega_j \hat{b}_j^{\dagger} \hat{b}_j+\frac{P_{1 j}}{\pi^2} L \ddot{h}(t) \sqrt{\frac{\hbar M}{\omega_j}}\left(\hat{b}_j^{\dagger}+\hat{b}_j\right),
\end{equation}

where we label the normal modes of increasing frequency to be $--,-, \mathrm{mid},+,++$.

For a detector with lumped element frequencies $\omega/2\pi = 800 \; \mathrm{Hz}$ range, we use $m_1 = 7000 \; \mathrm{kg}$, $\eta = 6 \times 10^{-5}$, which gives an end mass of $m_5 = 9.3 \times 10^{-14} \; \mathrm{kg}$. These parameters ensure that the coherent state amplitudes of each of the normal modes are such that there is approximately a $1 - \frac{1}{e}$ probability that one of the normal modes will transition to an excited state due to the absorption of a single graviton.

Similiarly, for a detector with lumped element frequencies $\omega/2\pi = 125 \; \mathrm{Hz}$, and for the same requirements of the transition probability, we use the effective masses of the first lumped element of the multi-mode system to be $m_1 = 20 \; \mathrm{kg}$, with $\eta = 6 \times 10^{-5}$, which requires the end mass to be $m_5 = 2.7 \times 10^{-16} \; \mathrm{kg}$.

Similarly for a detector with lumped element frequencies $\omega/2\pi = 2410 \; \mathrm{Hz}$ range, and first lumped element mass $m_1 = 2000 \; \mathrm{kg}$, with $\eta = 6 \times 10^{-5}$, which requires the end mass to be $m_5 = 2.6 \times 10^{-14} \; \mathrm{kg}$.

\bibliography{refs}

\end{document}